# Quantifying the reactivity of isolated Li$_x$Si domains in Si anodes using operando NMR

*Evelyna Wang\*, Marco-Tulio F. Rodrigues, Baris Key\**

AUTHOR ADDRESS

Chemical Sciences and Engineering Division

Argonne National Laboratory, Lemont, Illinois 60439.

AUTHOR INFORMATION

**Corresponding Author**

\*Evelyna Wang, evelyna.wang@anl.gov

\*Baris Key, bkey@anl.gov

## ABSTRACT

The use of Si anodes can greatly improve the energy density of Li-ion batteries. However, understanding and mitigation of calendar aging remains a barrier to commercialization. In this short report, we utilize operando Nuclear Magnetic Resonance (NMR) spectroscopy to detect and quantify lithium silicides ($Li_xSi$) as they form and react within Si anodes in pouch cells during calendar aging. We provide direct experimental evidence of complex aging phenomena in the Si anodes, including both SEI growth and dissolution during storage. Formation of electrochemically isolated $Li_xSi$ is also observed, as indicated by the partial persistence of highly lithiated phases after the cell is discharged. Remarkably, we show that these isolated domains can themselves self-discharge over time, suggesting that their detection can be challenging in post-mortem studies. Finally, we show that aging outcomes depend heavily on the type of silicon particles contained within the electrode, and that certain surface coatings can help decrease the reactivity between lithium silicides and the electrolyte.

**TOC GRAPHICS**

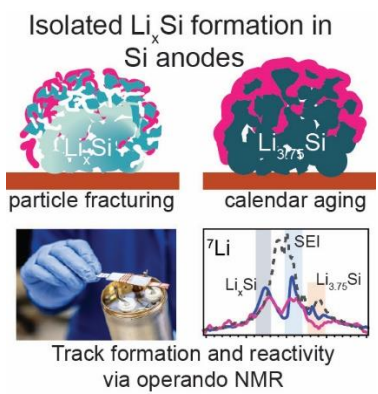

Lithium-ion batteries (LiBs) are ubiquitous in today's world of portable electronics and electric vehicles. In commercial LiBs, typical active materials for the cathode and anode are transition metal oxides and graphite, respectively.[1,2] Although substitution of graphite with silicon can increase the energy density of the cell,[3–5] Si anodes carry many inherent challenges.[4–7] The lithium silicides that form during lithation of Si are highly reactive Zintl phases, posing cycle and calendar life challenges.[8–10] In addition, Si anodes undergo severe volume changes during lithiation and delithiation, resulting in mechanical degradation.[6,11] The combination of these two properties of Si results in an unstable and non-passivating solid-electrolyte interphase (SEI) layer, when compared to graphite.[11–13] The challenges with Si anodes necessitate the use of non-destructive characterization methods to observe the reactive lithium silicide ($Li_xSi$) species and SEI instability. Furthermore, full-cell testing of Si anodes is required to understand long-term cycle and calendar life aging behavior.[14,15] As such, we utilized non-destructive, operando Nuclear Magnetic Resonance (NMR) spectroscopy to probe the reactivity of lithium silicides during calendar aging of Si anodes in full-cells.

Calendar aging studies are relevant to practical battery applications, as they establish an upper bound for the lifetime of the cell.[14] Extended storage of cells at high states of charge (SOCs) leads to self-discharge and permanent capacity fade, due to SEI growth and losses of Si active material.[15–18] Calendar aging in Si/graphite composite electrodes have been previously shown to be worse than calendar aging in graphite only electrodes in terms of resulting capacity fade.[14] This was attributed largely to the less passivating nature of the SEI on Si active materials. The SEI instability on Si anodes was further highlighted by Stetson et al. whereby decreases in SEI resistivity and thickness as well as increases in anode potential were observed during rest.[19] McBrayer et al. probed the SEI instability using Scanning Electrochemical Microscopy and highlighted chemical modes of failure, rather than mechanical, as major challenges in Si anodes.[20] Furthermore, common additives to stabilize the SEI and improve cycle and calendar aging stability in conventional Li-ion batteries may not be viable for Si anodes.[21]

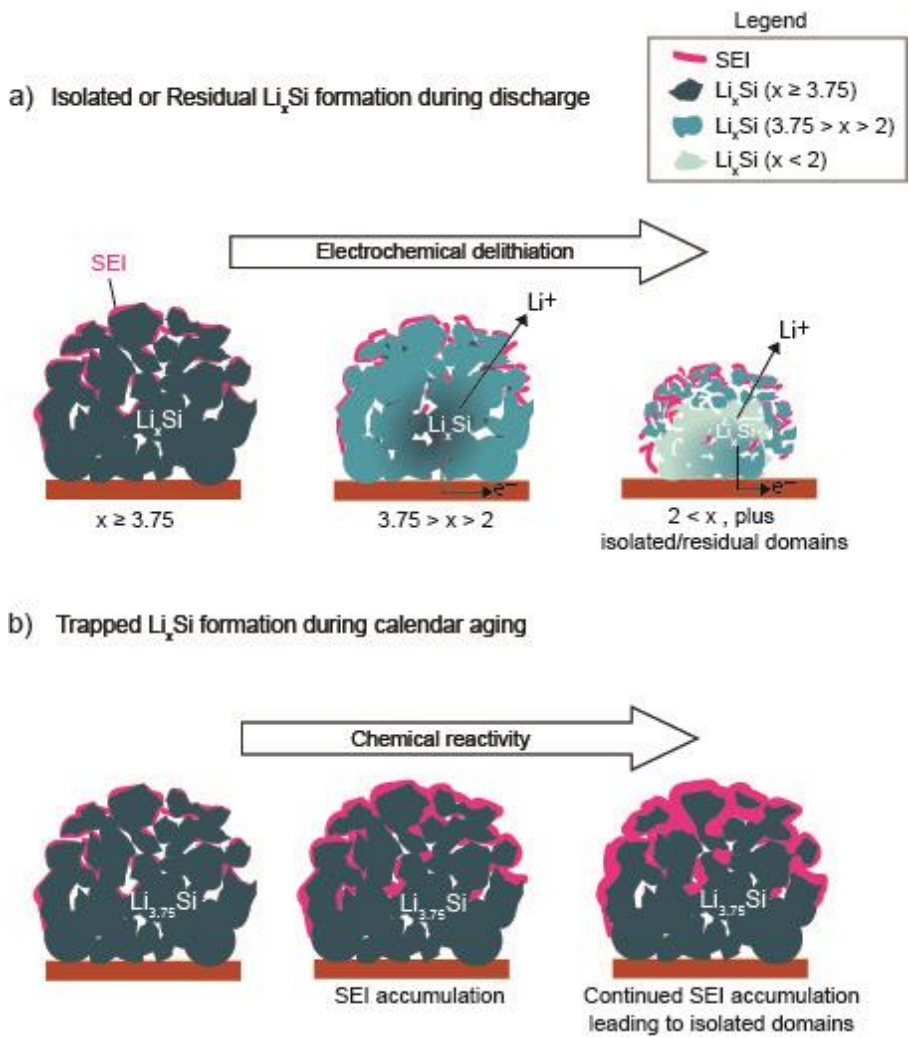

**Figure 1**. Two potential mechanisms for the formation of isolated Li$_x$Si phases within Si anodes: a) during electrochemical delithiation, the fracturing of particles results in isolation of lithiated Si domains, and b) during calendar aging, the chemical reactivity of Li$_x$Si accumulates SEI species that isolate lithiated Si domains

Si active material may become electrically isolated either from mechanical degradation (i.e. cracking and pore formation) or from the blocking of Li$^+$ transport due to SEI formation (i.e. chemical/electrochemical reactivity).[15,18,22,23] Figure 1 illustrates two mechanisms for Si active material isolation: (a) during electrochemical delithiation (discharge in a full-cell), the volume contraction of Si particles leads to fracturing and isolation of partially delithiated phases, and (b) during calendar aging at high SOC, the chemical reactivity of highly lithiated phases and non-passivating nature of the SEI can result in continuous SEI formation, leading to the trapping of highly lithiated phases.[17,18] The accumulation of isolated phases contributes to loss of Li inventory (LLI) and loss of active Si material (LAM). Quinn et al. developed advanced microscopy techniques to observe intra-particle Si fragmentation and SEI growth that led to the LAM, as well as trapping of previously mobile Li$^+$ ions via SEI buildup that lead to LLI.[24] In this work, we refer to the isolated and lithiated Si domains as "trapped Li$_x$Si". Using operando Nuclear Magnetic Resonance (NMR) spectroscopy, we investigated the formation and the subsequent reactivity of these trapped Li$_x$Si species.

NMR spectroscopy is an ideal tool for investigating local chemical environments within samples, and has been successfully utilized in Si anodes.[25,26] NMR has previously been used to identify the step-wise formation of lithium silicides,[27–29] and the chemical shifts of compounds with varying degrees of lithium content, ranging from Li$_x$Si (x < 2) to Li$_{3.75}$Si (or Li$_{15}$Si$_4$), have been well characterized. Building upon previous NMR studies of Si anodes, we have developed operando $^7$Li NMR methods to probe the direct formation and evolution of reactive lithium silicides as well as their chemical and electrochemical reactivity during calendar and cycle aging.[27–31] This operando NMR method can probe the Si anodes within full-cells assembled via large-scale fabrication methods comparable to commercial pouch cells.[31]

Here, we present a short report using the operando $^7$Li NMR method to identify the formation of trapped Li$_x$Si species in full-cells comprising of Si anodes and NMC cathodes. First, we observed the calendar aging behavior of milled Si anodes at high states of charge and quantified the chemical reactivity of lithium silicides within the anode. Following discharge of the cells, we observed the formation of trapped Li$_x$Si species; we were able to distinguish between isolated species that had formed during the calendar aging or during the discharge. Furthermore, we observed that during calendar aging at low SOC, the isolated Li$_x$Si were chemically reactive. Finally, we compared the calendar aging behavior and reactivity of formed Li$_x$Si phases at moderate states of charge as well as tested alternative nano-particulate Si anodes prepared via plasma-enhanced chemical vapor deposition.[32]

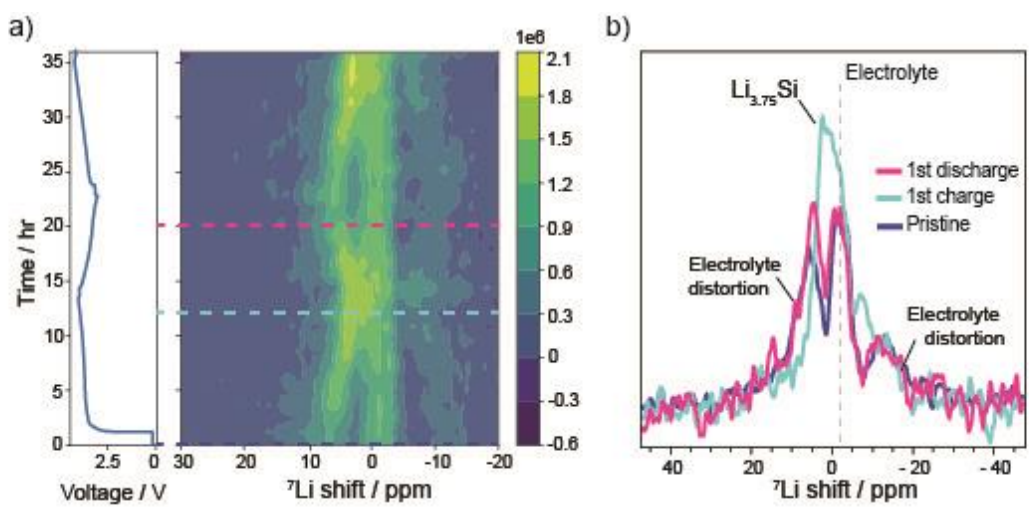

**Figure 2.** (a) The voltage versus time is plotted for a full-cell comprising milled Si anode|NMC811 along with the contour plot showing $^7$Li NMR spectra collected in operando during cell charge and discharge. (b) The $^7$Li NMR spectra corresponding to the charged cell (lithiated Si anode) is plotted with the discharged (delithiated Si anode) and the pristine spectra

Figure 2a exhibits the voltage versus time data for the initial cycles of a NMC vs Si cell, along with the operando NMR contour plot. The anode in the cells contains ball-milled Si, resulting in particles with ≥ 260 nm of diameter.[33] We observe the formation of a peak between 2 and 4 ppm during cell charge (lithiation of Si) and its subsequent disappearance during discharge (delithiation of Si). We attribute the peak that initially forms around 4 ppm to Li in extended Si clusters ($Li_xSi$, x < 2). The $Li_xSi$ peak then shifts toward 2 ppm due to increasing lithium content. At sufficiently high states of Si lithiation (at ~3.95 V in the full-cell), the peak at around −7 ppm forms, corresponding to Li-rich phases $Li_{15+x}Si_4$ ($Li_xSi$, x ≥ 3.75).[27,28] This lithiation behavior is characteristic of crystalline Si anodes.[29] The peak for highly lithiated $Li_xSi$ (x ≥ 3.75) is the first to disappear at the start of discharge, followed by a decrease in peak intensity for the peak at 2 ppm and a shift toward 4 ppm.

Specific $^7Li$ NMR spectra for the pristine cell, end of first charge, and end of first discharge are shown in Figure 2b to better illustrate the $Li_xSi$ peaks. For the pristine spectra, only the Li environment from the electrolyte is expected to give rise to signal; the Li environments in the lithiated NMC 811 with fast relaxation times were not directly measured with the acquisition parameters used.[34] However, we observe a distortion in the bulk electrolyte signal for the pristine NMR spectra, which results in line broadening, asymmetry, and two additional peaks on either side of the bulk electrolyte peak at approximately 7 ppm and −12 ppm. This distortion is caused by the overall static nature of the measurement and any local non-uniformity experienced by the Li ions in the electrolyte as well as the bulk magnetic susceptibility (BMS) effect.[35] These effects and the analysis required to obtain the underlying structural information are discussed further in our previous work and in the experimental section.[31]

When the cell is charged, the moderately lithitiated Li$_x$Si (2 < x < 3.75), and highly lithiated Li$_x$Si (x ≥ 3.75) peaks at 2 ppm and −7 ppm are clearly observed and the distorted electrolyte peak at 7ppm is also removed (due to the lithiation of the NMC cathode). After the first discharge, the Li$_x$Si (x ≥ 3.75) peak is largely removed. The spectra after the first discharge exhibits residual signal around 4 ppm, indicating that some Li$_x$Si (x < 2) species remain and that the Si was not fully delithiated. These species are residual due to isolation (i.e. fracturing, entrapment in SEI etc.) or insufficient delithiation potential at the anode during discharge; the slopy voltage profile of Si at low states of lithiation can cause the anode to hold non-negligible amounts of Li$^+$ even when cells are nominally fully discharged.[36]

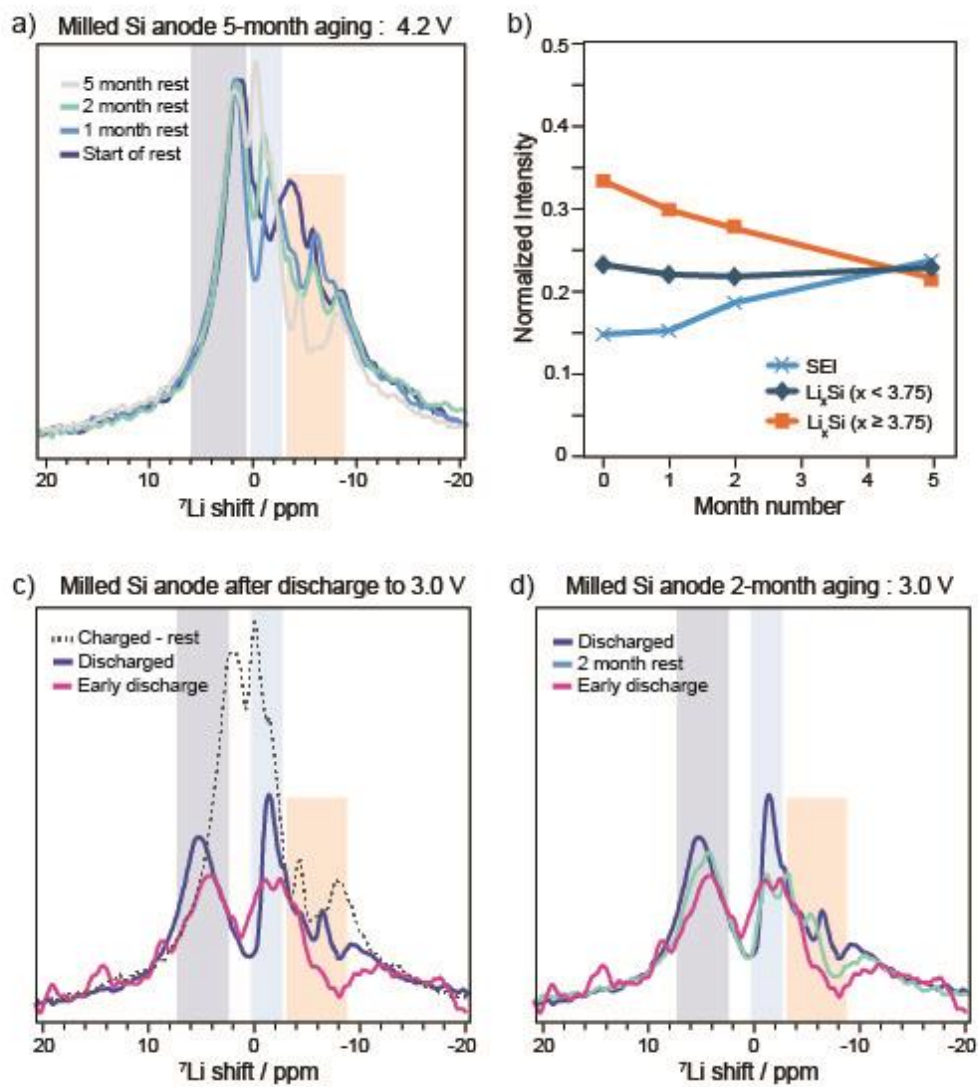

**Figure 3.** (a) The $^7$Li NMR spectra for calendar aging of a full-cell with a milled Si anode charged to 4.2V and left to rest. (b) The signal integrations for regions corresponding to the SEI, $Li_xSi$ (x < 2), and $Li_xSi$ (x ≥ 3.75) plotted against calendar aging time in months. Signal integrations were normalized to the values at the start of rest to get the fractional intensity. (c) The $^7$Li NMR spectra for the same calendar aged Milled Si full-cell after discharge to 3.0 V, comparing the formation of isolated or residual $Li_xSi$ domains. (d) The $^7$Li NMR spectra for the same cell that was then subject to calendar aging for at low SOC showing the chemical reactivity of the isolated $Li_xSi$ domains.

Next, we investigated the formation of isolated and trapped Li$_x$Si domains during calendar aging at high SOC. In Figure 3a, we show the calendar aging behavior for the milled Si anode full cell at high SOC; the cell was charged to 4.2V and left to rest at room temperature (~25 °C) for 5 months. In the fully charged $^7$Li NMR spectra, we observe Li$_x$Si (x ≥ 3.75) species with intensity at around −4 ppm to −8 ppm, highlighted in the orange region. We also observe moderately lithiated phases, Li$_x$Si (2 < x < 3.75), with intensity around 2-6 ppm (Figure 2a, dark blue highlight). Signal contribution from SEI species (such as LiF or Li carbonates) and electrolyte salt (LiPF$_6$) are included in the region around −2 ppm to 0 ppm (Figure 3a, light blue highlight). Figure 3b shows the fractional intensities during calendar aging for the regions [0 to −2 ppm], [2 to 6 ppm], and [−4 to −8 ppm]. The signal integrations over the chemical shift regions were normalized to the integration values at the start of rest to give the fractional intensities plotted in Figure 3b. We observe a consistent decrease in the highly lithiated species during rest and an increase in diamagnetic species, suggesting the chemical reactivity of the highly lithiated silicides to form SEI. The less lithiated Li$_x$Si phases show less reactivity during the calendar aging, as the fractional signal intensity only decreased slightly.

Following the 5-month calendar aging at high SOC, the cell was subsequently discharged to 3.0 V. Figure 3c exhibits the final $^7$Li NMR spectra for the charged cell at the end of the 5-month rest, shown in the dotted gray trace, and the subsequent discharged spectra, shown in the dark blue trace. The majority of the lithiated Li$_x$Si (x ≥ 3.75) and Li$_x$Si (x < 3.75) species are removed at the end of discharge. However, compared to the $^7$Li NMR spectra for the 3$^{rd}$ discharge shown in the pink trace, there is more signal intensity for both Li$_x$Si (x ≥ 3.75) and Li$_x$Si (x < 3.75) in the discharge after calendar aging at high SOC. The presence of the highly lithiated Li$_x$Si (x ≥ 3.75) even after discharge suggests that these species became isolated during the calendar aging and were subsequently unable to delithiate, as summarized in Figure 1a.[17,18] In contrast, the less lithiated phases form during the discharge step due to either particle fracturing or delithiation overpotentials. In addition, we observe a large increase in diamagnetic species including SEI for the discharge after calendar aging at high SOC compared to an earlier discharge.

The cell was then left to calendar age in the discharged state for 2 months, after which another NMR spectrum was collected (Figure 3d, cyan trace). There is a large decrease in the signal intensity for the diamagnetic Li region, which can be explained by a dissolution of the Li-bearing SEI species during calendar aging. The SEI dissolution indicates that it is unstable and can expose the residual $Li_xSi$ phases to further chemical reactions with the electrolyte. Indeed, we also observe a decrease in the signal intensity around −8 ppm as well as a decrease in signal intensity around 4 ppm during rest, indicating chemical reactivity of both the highly lithiated and trapped $Li_xSi$ (x ≥ 3.75) species and the less lithiated silicides.

Approximately 30% of the fractional signal intensity for the $Li_xSi$ (x ≥ 3.75) species has disappeared after the 5-month calendar rest. This corresponds well to the decrease in discharge capacity after the rest at high SOC, whereby only 75% of the charge capacity was obtained on the subsequent discharge following the 5-month calendar aging. This capacity loss with calendar aging is exacerbated in comparison to other literature studies due to the lack of external pressure applied in our cells.[37] The effects of pressure on cell capacity are however currently outside the scope of our study as we focus on quantifying the reactivity of the lithium silicides.

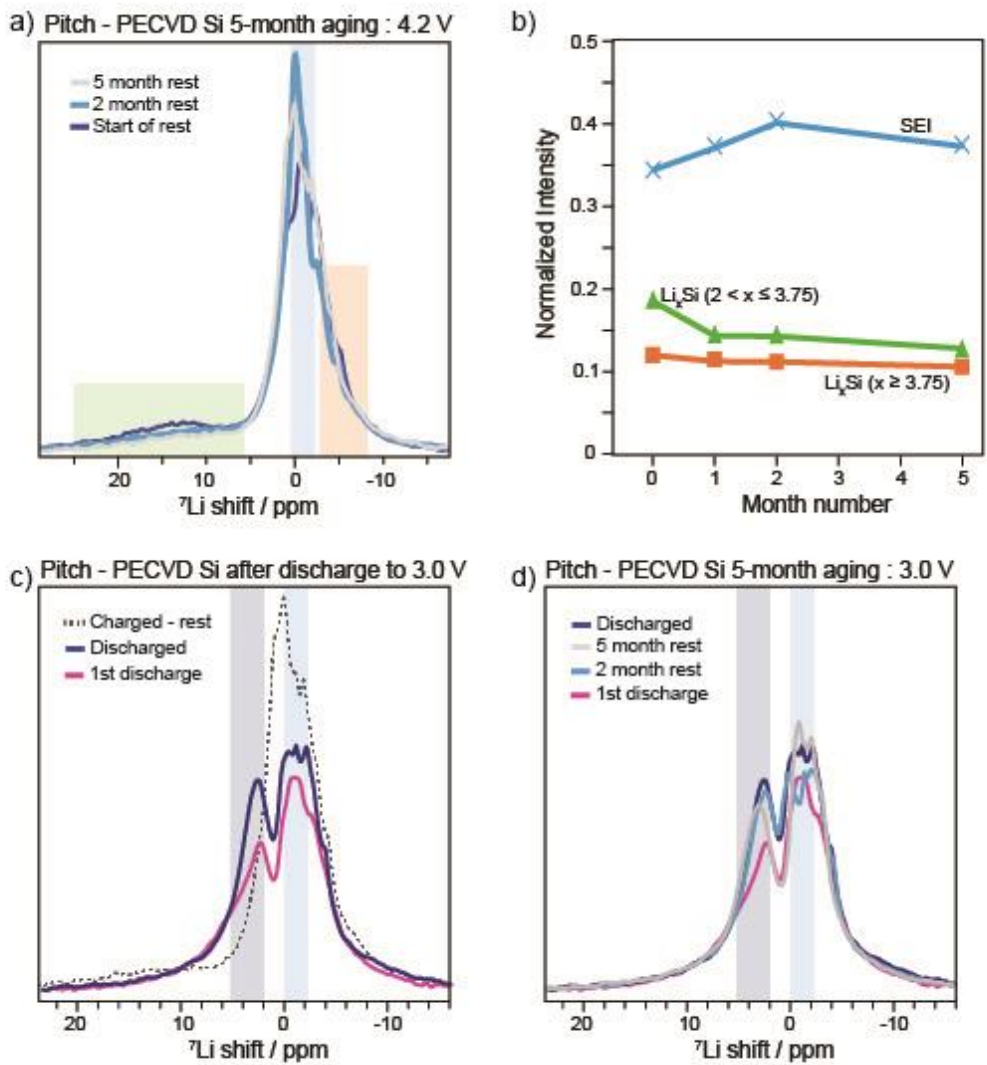

**Figure 4.** (a) The $^7$Li NMR spectra for calendar aging for a full-cell with a pitch-coated PECVD charged to 4.2V and left to rest. (b) The fractional signal integrations for regions corresponding to the SEI, Li$_x$Si ($2 < x \leq 3.75$), and Li$_x$Si ($x \geq 3.75$) plotted against calendar aging time in months. Signal integrations were normalized to the values at the start of rest to get the fractional intensity (c) The $^7$Li NMR spectra for the same calendar aged PECVD Si full-cell after it was discharged to 3.0V. (d) The $^7$Li NMR spectra for the same discharged PECVD Si full cell that was then subject to calendar aging at this low SOC.

Finally, we also compared the calendar aging performance of a different Si anode material in full cells using our operando NMR methodology. The $^7$Li NMR spectra collected during calendar aging at high SOC and low SOC for PECVD-grown nano-scaled Si coated with pitch carbon are shown in Figure 4. This pitch-coated PECVD-grown Si is comprised of ≤10-nm Si particles.[38] As such, we observed a slightly different lithiation mechanism in the nanoparticulate PECVD Si compared to the milled Si; here we observe the formation of Li in small Si clusters in addition to highly lithiated phases.[28,31] In the charged PECVD Si full-cell shown in Figure 4a, we observe Li$_x$Si (2 < x ≤ 3.75) species, such as Li$_{13}$Si$_4$ or amorphous Li$_{15}$Si$_4$, at ~15 ppm. These species decrease during calendar aging due to the chemical reactions at rest. Again, similar to the milled Si, the calendar aging at rest results in chemical degradation of lithiated silicides, leading to self-discharge and capacity loss; the discharge capacity following the calendar aging at high SOC was 84% of the charge capacity prior to rest, suggesting 16% capacity was lost due to the self-discharge. Compared to the milled Si anode, the lithium silicides in the pitch-coated PECVD Si shows less chemical reactivity even while resting at high SOC; the evolution of the fractional intensities shown in Figure 4b were more stable compared to the milled Si in Figure 3b. This difference in reactivity may be attributed to the carbon coating which has been reported to help stabilize the reactive species.[32,39,40]

Next, we compared the $^7$Li NMR spectra at the end of the high SOC calendar aging for the PECVD Si full-cell (dotted trace, Figure 4c) and its subsequent discharge to 3.0V (blue trace, Figure 4c). We do not observe the presence of highly lithiated silicides, indicating that during calendar aging at high SOC, there was minimal isolation of such Li-rich phases due to chemical reactivity. This is in stark contrast to the milled Si full-cell, where isolated highly lithiated silicides were clearly observed at the end of discharge following the calendar aging at high SOC, as shown in Figure 3c. The milled Si anode with larger Si particles (~260nm) is more prone to fracturing and formation of trapped $Li_xSi$ compared to the nanoparticulate PECVD Si (≤10nm). Nevertheless, after extended storage at high SOC, the discharged and delithiated PECVD Si anode still shows accumulation of some $Li_xSi$ (x < 2) phases and SEI species as shown in Figure 4c; comparing the calendar-aged and discharged PECVD Si anode (dark blue trace) to the early discharge (pink trace), we can observe the accumulated $Li_xSi$ (x < 2) highlighted in the dark blue region and the accumulated SEI highlighted in the light blue region.

After discharging to 3.0 V, the PECVD Si full-cell was then left to calendar age at low SOC (Figure 4d). We observed a slight chemical reactivity of the residual $Li_xSi$ (x < 2) phases during calendar aging, as suggested by a small decrease in the signal intensity at 3ppm. These less lithiated silicides are much more stable than the trapped highly lithiated $Li_xSi$ phases observed in the milled Si (Figure 3d). In the PECVD Si full-cell, there was also variation in the 0 ppm region corresponding to SEI species during the low SOC calendar aging. This suggests that both dissolution of the SEI and accumulation of SEI, due to chemical reactions of the residual $Li_xSi$, may occur during low SOC calendar aging for this material.

In conclusion, this work highlights the reactivity of lithium silicides, as probed via operando NMR. We observed trapped lithium silicides that form in the first cycle and continue to accumulate with additional aging. Interestingly, these trapped domains remain chemically reactive and will self-discharge over time, likely due to chemical reactions with the electrolyte. An important consequence of this continued reactivity is that attempts to identify and quantify these domains post-mortem may underestimate their true abundance. The quantification of these trapped $Li_xSi$ species and reporting of their chemical reactivity have not been well discussed prior to this study. Our operando technique also revealed intricate phenomena related to SEI growth. Our data shows clear indication of self-discharge of Si anodes during aging at high states of lithiation, with concomitant increase in the signal related to SEI species. However, we also present evidence of SEI dissolution occurring during extended storage of the cell, even in the discharged state. Furthermore, self-discharge and SEI growth occurred more prominently when $Li_xSi$ phases with a higher $Li^+$ content (x ≥ 3.75) were present, suggesting that calendar life could be improved by constraining Si utilization. Finally, we determined that aging outcomes were highly material-dependent, with the silicides in a nano-particulate PECVD Si (≤10 nm) coated with pitch carbon being less reactive than in milled Si (~260 nm) and less prone to isolation. This latter observation shows that surface coatings may help mitigate the intrinsic reactivity expected for smaller particles.

EXPERIMENTAL

Cell assembly: All cells were assembled in a humidity controlled dry room. Milled Si anode and NMC 811 cathode laminates were prepared by the Cell Analysis, Modeling, and Prototyping (CAMP) Facility. Milled Si anodes were obtained from collaborators at Oak Ridge National Laboratory and produced as described in previous works.[33,41] In brief, Si boules underwent high-energy milling to produce Si particles ~260 nm in size. PECVD anode laminates were provided by collaborators from NREL and produced as described in their previous work.[38] For the operando NMR pouch cells, the electrodes were cut (2.2 cm x 1 cm anode and 1.9 cm x 1 cm cathode) and welded onto current collector tabs (Cu was used for the Si anode and Al was used for the cathode). The separator used was Celgard 2500 (25 μm microporous monolayer PP membrane). Electrodes and separator were sealed into an aluminum-polymer (polypropylene) laminate pouch. The cells were filled with 100 μL of 1.2 M $LiPF_6$ in ethyl methyl carbonate (70% wt.) and ethylene carbonate (30% wt.) electrolyte with 3% fluoroethylene carbonate additive. The filled cells were then sealed. The N/P ratio was ~0.8; overcharge is averted due to Li+ consumption during the initial SEI formation.

Calendar aging at high states of charge: The cell was subjected to 10x C/10 cycles, then charged to 4.2V and left to calendar age at room temperature and OCV.

Calendar aging at low states of charge: After calendar aging at high SOC, the cells were subsequently discharged to 3.0 V and then left to calendar age at room temperature and OCV.

Operando $^7$Li NMR: All measurements were performed on a 300 MHz (7.04 T) Bruker Avance III Spectrometer using a static probe. A custom Cu solenoid coil was prepared, maximizing the fill-factor of the pouch cells. For the $^7$Li measurements, a one pulse sequence was used with 1 s recycle delays. The measurements were referenced to a 1 M LiCl aqueous sample sealed in plastic. Care was taken to place the cell in the same position for each measurement and minimize effect of BMS in changing the line shape when comparing between measurements. Local non-uniformities that affect the BMS include electrolyte wetting onto electrode surfaces, current collectors, or edges of the pouch material, which result in different spin relaxation and peak broadening. In addition, paramagnetic species can alter the observed $^7$Li NMR electrolyte chemical shift; Li ions within the electrolyte wetted onto the paramagnetic NMC811 cathode surface will be shifted compared to freely moving Li ions in electrolyte. Changes in the transition metal oxidation states as the NMC811 cathode is charged or discharged will also affect this shift.

AUTHOR INFORMATION

**Notes**

The authors declare no competing financial interests.

ACKNOWLEDGMENTS

This research was supported by the U.S. Department of Energy's Vehicle Technologies Office under the Silicon Consortium Project, directed by Brian Cunningham, Thomas Do, Nicolas Eidson and Carine Steinway, and managed by Anthony Burrell. The submitted manuscript has been created by UChicago Argonne, LLC, Operator of Argonne National Laboratory ("Argonne"). Argonne, a U.S. Department of Energy Office of Science laboratory, is operated under Contract No. DE-AC02-06CH11357. The U.S. Government retains for itself, and others acting on its behalf, a paid-up nonexclusive, irrevocable worldwide license in said article to reproduce, prepare derivative works, distribute copies to the public, and perform publicly and display publicly, by or on behalf of the Government. The cells in this article were fabricated at Argonne's Cell Analysis, Modeling, and Prototyping (CAMP) Facility. We also thank Nathan Neale and NREL team for providing the pitch coated PECVD grown silicon nanoparticles.